\documentclass{article}

\usepackage{arxiv}

\usepackage[utf8]{inputenc} 
\usepackage[T1]{fontenc}    
\usepackage{hyperref}       
\usepackage{url}            
\usepackage{booktabs}       
\usepackage{amsfonts}       
\usepackage{nicefrac}       
\usepackage{microtype}      

\usepackage{fullpage}
\usepackage{amsmath} \usepackage{amssymb}
\usepackage{bm} \usepackage[rgb,dvips]{xcolor}
\usepackage{listings} \usepackage{graphicx}
\usepackage{longtable}
\usepackage[english]{babel}

\title{Latent Space Explorations of Singing Voice Synthesis using DDSP}

\author{Juan Alonso and Cumhur Erkut\\
  Department of Architecture, Design, and Media Technology\\
  Aalborg University Copenhagen\\
  Denmark\\
  \url{https://www.smc.aau.dk/}}

\begin{document}
\maketitle
\begin{abstract}
Machine learning based singing voice models require large datasets and lengthy training times. In this work we present a lightweight architecture, based on the Differentiable Digital Signal Processing (DDSP) library, that is able to output song-like utterances conditioned only on pitch and amplitude, after twelve hours of training using small datasets of unprocessed audio. The results are promising, as both the melody and the singer's voice are recognizable. In addition, we present two zero-configuration tools to train new models and experiment with them. Currently we are exploring the latent space representation, which is included in the DDSP library, but not in the original DDSP examples. Our results indicate that the latent space improves both the identification of the singer as well as the comprehension of the lyrics. Our code is available at \url{https://github.com/juanalonso/DDSP-singing-experiments} with links to the zero-configuration notebooks, and our sound examples are at \url{https://juanalonso.github.io/DDSP-singing-experiments/}. 
\end{abstract}

\section{Introduction}\label{ch:ch1label}
The human voice is one of the oldest musical instruments \cite{fruhholz2018oxford}. In the era before Deep Learning \cite{lecun2015deep}, high-quality singing synthesis was carried out either by the spectral models \cite{serra1990spectral}, based on perception, or the physical models \cite{Cook96:CMJ}, based on production and articulation. Combining the spectral models with deep learning, the Differentiable Digital Signal Processing \cite{engel2020ddsp} (DDSP) library by Google's Magenta team became a powerful toolkit for audio-related machine learning. The first published examples of DDSP were focused on timbre transfer from monophonic instruments.

In this paper we present the DDSP architecture to a more complex, expressive instrument: the human vocal apparatus, and check the suitability of the DDSP for singing voice synthesis. By constructing small size databases, experimenting with the model parameters and configurations, and by training the resulting models only for about twelve hours, we obtain singing-like outputs, which clearly resemble to original singers / speakers. However, the lyrics are incomprehensible because we don't have a language model. Then, by conditioning the model on the MFCC of the original audio and creating a latent space, the results improve, showing promising intelligibility. Our contribution also includes extensive documentation of the library and two zero-config notebooks for experimentation.

This paper is organized as follows. In Sec~\ref{ch:ch2label}, we introduce the context of neural singing synthesis. Next we provide a detailed look at the DDSP architecture for timbre transfer. In Sec.~\ref{ch:ch4label}, we introduce our experiments together with the data sets and model configurations, and the improved results we have obtained by adding our latent space explorations. In the next section, we discuss our observations. We finally draw conclusions and indicate possible areas of further research. 

\section{Background}\label{ch:ch2label}
In neural singing synthesis, the Deep Neural Network (DNN) receives a list of pitches and the lyrics as input, and outputs a signal modeling a specific voice. It is a problem more complex than speech synthesis because of more diverse sets of pitches and intensities, different vowel durations, etc, but still closely related to it.

Gómez et al. \cite{gomez2018deep} revised many data-driven deep learning models for singing synthesis. A thoughtful remark in that paper is that the black-box characteristics of deep learning models make it very difficult to gain knowledge related to the acoustics and expressiveness of singing. Even if deep learning techniques are general and can learn from almost any arbitrary corpus, it is necessary to advocate for explainable models to break the black-box paradigm.

The DDSP library is the latest set of tools released by Google's Magenta; a team dedicated to exploring the role of machine learning as a creative tool. DDSP is set to bring explainability and modularity in neural audio synthesis \cite{engel2020ddsp}. The idea behind DDSP is "\emph{to combine the interpretable structure of classical DSP elements (such as filters, oscillators, reverberation, envelopes...) with the expressivity of deep learning.}"\footnote{\url{https://magenta.tensorflow.org/ddsp}} To avoid a common misunderstanding, DDSP is not an architecture \textit{per se}, but a set of signal-processing tools that can be incorporated into modern automatic differentiation software. Many examples of DDSP relate to signing input, therefore we provide an overview in this section.

Several papers extended the DDSP specifically for speech and signing synthesis \cite{hutter2020timbre,fabbro2020speech,jonason2020control}. In \cite{hutter2020timbre}, the model is conditioned on the phoneme level using an Automatic Speech Recognition (ASR) system to extract the phonemes of the training set and
use them as additional conditioning data. In \cite{fabbro2020speech}, the authors
synthesize spoken speech using the DDSP architecture with a model
conditioned on mel-spectrograms, instead of using raw audio. The loss
function is also adapted to use mel-spectrograms trying to mimic the way
human perception works. In \cite{jonason2020control}, the authors propose using MIDI
data modified by a LSTM network, to obtain continuous pitch and loudness
contours that will be fed into the DDSP architecture.

\subsection{DDSP Overview}\label{DDSP overview}
Phase alignment poses a problem when generating audio in deep learning; that problem is present when frames are used, either in the time- or the frequency-domain. An autoregressive model does not present this problem, but instead is harder to train due to the amount of data needed, and the interplay between audio perception, wave shape and loss (two wave shapes with very different losses can be perceived as sounding exactly the same).

One possible solution is the use of audio oscillators or vocoders that perform analysis and synthesis. In analysis, interpretable parameters such as f0 or loudness are extracted, and in synthesis the generative algorithm uses these parameters to construct the synthetic sound. The problem with vocoders is that they are difficult to incorporate into a machine learning workflow, because their parameters are not differentiable. Therefore, they cannot backpropagate during the training phase. DDSP solves this problem by rewriting a series of classical Digital Signal Processing (DSP) modules as feedforward functions allowing efficient implementation in GPUs.

\section{DDSP and timbre transfer}\label{ddsp-and-timbre-transfer}

\begin{figure}
  \hypertarget{fig:fig_timbre_transfer_architecture}{%
    \centering \includegraphics[width=\columnwidth]{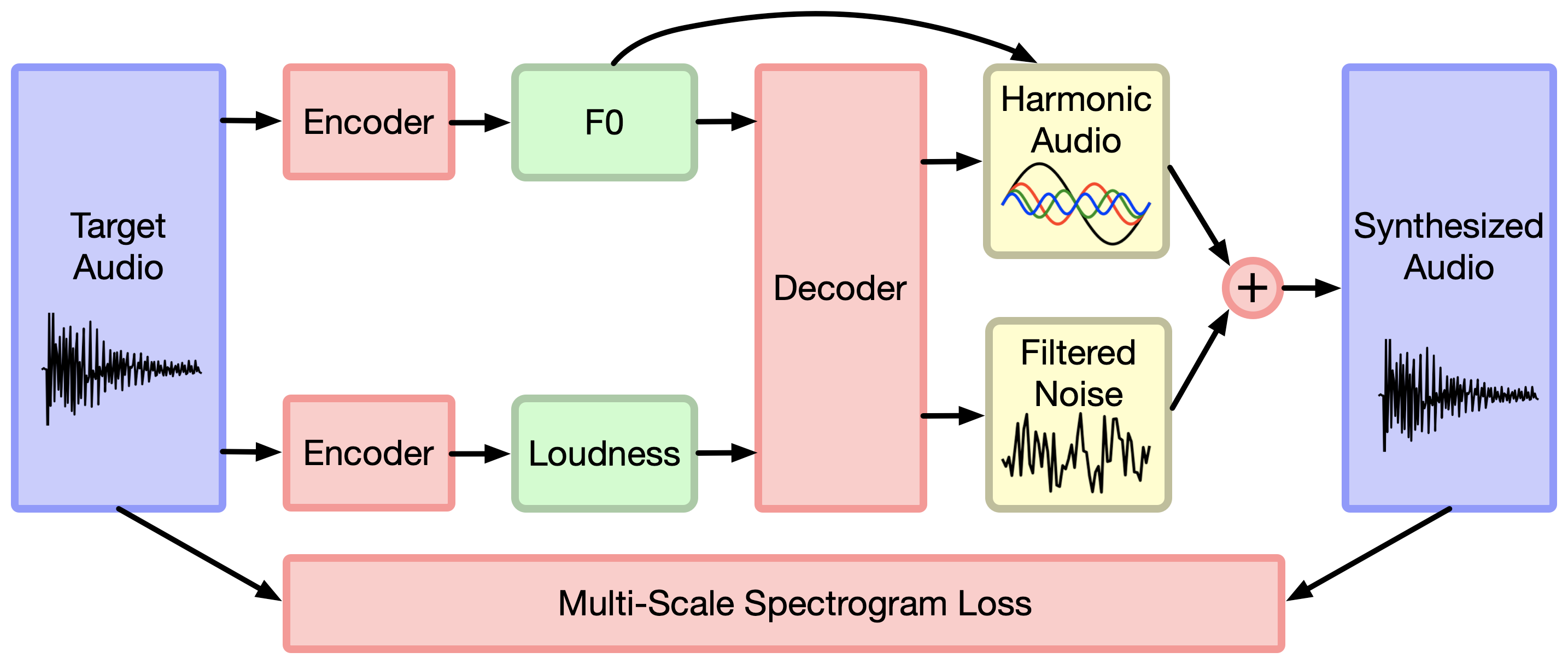}
    \caption{Timbre transfer architecture used in this work. Adapted from
\cite{engel2020ddsp}.}\label{fig:fig_timbre_transfer_architecture}
}
\end{figure}

The architecture used in this work is based on the "solo\_instrument" setup proposed for timbre transfer in \cite{engel2020ddsp} and shown in Fig~\ref{fig:fig_timbre_transfer_architecture}. The network is presented an audio file. f0 and loudness are extracted and fed to the decoder, which produces the parameters for a harmonic synthesizer and for a subtractive synthesizer. The audio from both synthesizers is combined and presented as the final output. During the training phase, the loss is computed using different resolution spectrograms from the original and the resulting signal. In the following sections, we will describe in detail all the modules used.

The {\bf encoder} transforms the incoming audio into latent vectors, in this case, two interpretable features: the fundamental frequency and the perceived loudness of the monophonic input audio. The DDSP library does not require a specific method to generate the \textbf{f0} and \textbf{loudness} vectors. The expected format is an array of float values, sampled at 250Hz.

\textbf{f0} can be generated synthetically, but the DDSP library includes examples using CREPE \cite{kim2018crepe} and DDSP-inv \cite{engel2020self}.

\begin{itemize}
\item \textbf{CREPE} is a monophonic pitch tracker model based on a deep convolutional neural network that operates on the time domain. The model input, a 1024-sample chunk of audio at 16kHz, is fed into a series of convolutional layers, a dense layer and finally an activation layer with 360 outputs, corresponding to pitches between 32.70Hz and 1975.5Hz (MIDI C1 to B6), where each output covers 20 cents of a semitone.

\item \textbf{DDSP-inv} is based on the DDSP architecture and uses two neural networks. The first one, the sinusoidal encoder, uses log mel-spectrograms to compute the sinusoidal amplitudes, sinusoidal frequencies and noise magnitudes of the audio. The second network converts this data into fundamental frequency, amplitude and harmonic distribution.
\end{itemize}

It is worth noting that the f0 latent vector is fed directly to the additive synthesizer. This allows to disentangle the fundamental frequency and facilitates the model to respond to frequencies unseen during the training and improves the expressiveness of the output.

\textbf{Loudness} can also be generated synthetically, but the DDSP library includes a utility function to compute the perceptual loudness of the audio, applying A-weighting curves to the power spectrogram.

\begin{figure}[tbph]{}
  \centering
  \includegraphics[width=0.75\linewidth]{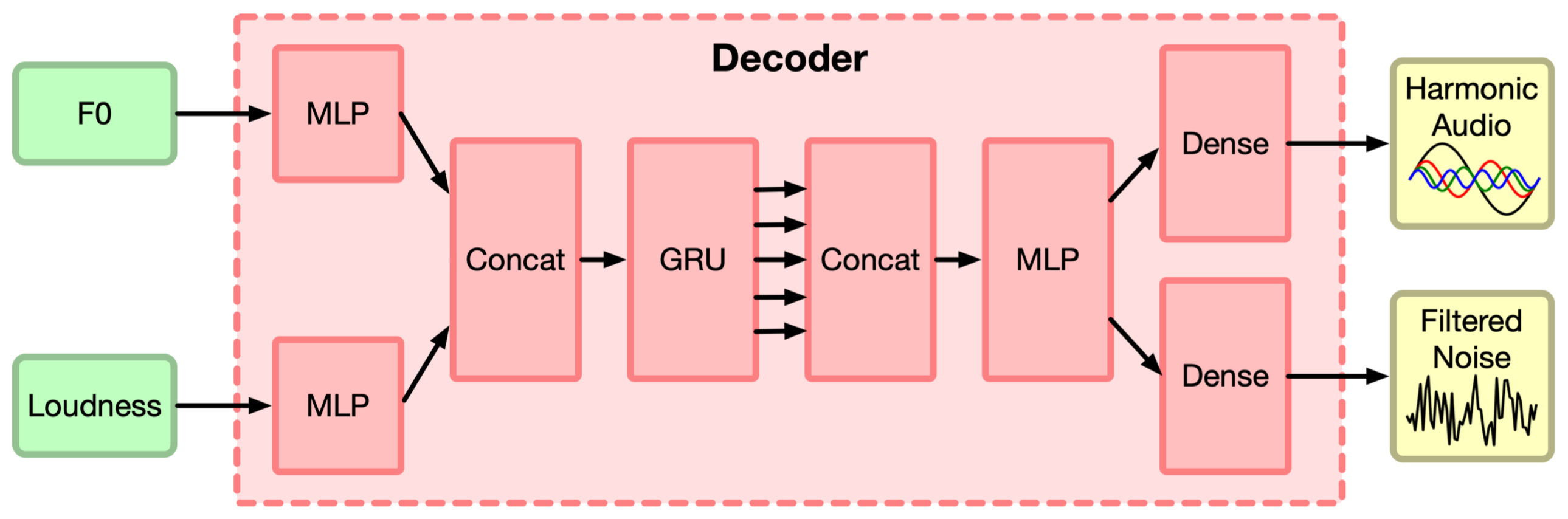}
  \includegraphics[width=0.75\linewidth]{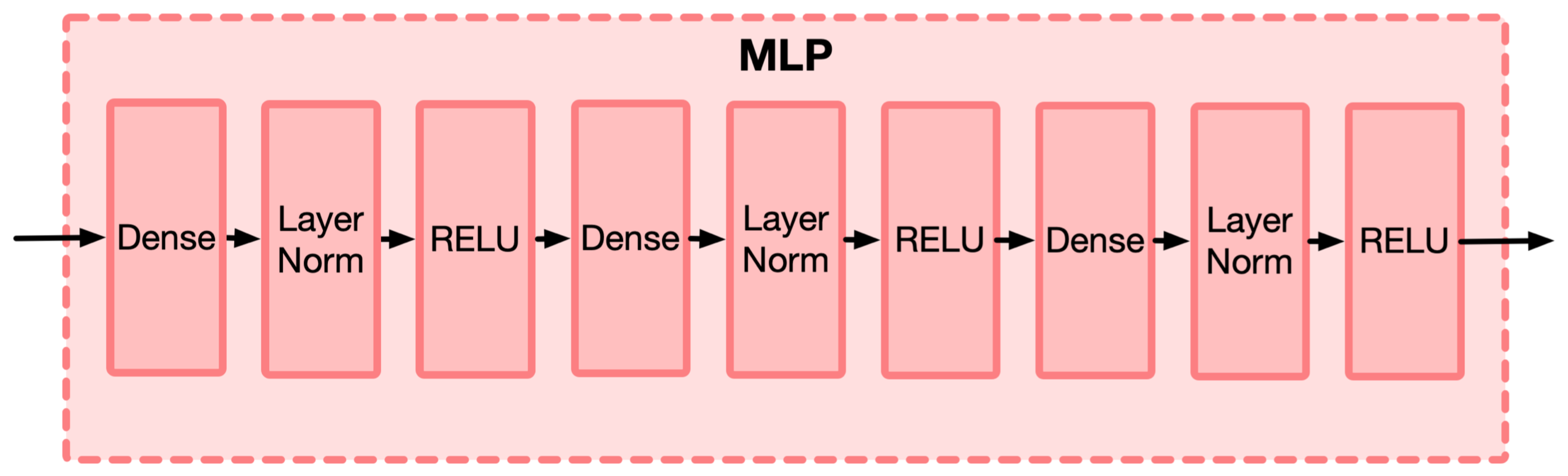}
  \caption{Decoder architecture (top) and MLP structure (bottom). Adapted from \cite{engel2020ddsp}.}
  \label{fig:fig_decoder}
\end{figure}

The \textbf{decoder} (Figure \ref{fig:fig_decoder}, top) is an RNN that receives the latent vectors (f0 and loudness) and outputs the control parameters required by the synthesizers: the amplitudes of the harmonics, and the transfer function for the FIR filter. The RNN is fairly generic, as the DDSP authors want to emphasize that the quality of the results comes from the DDSP modules, not from the complexity of the neural network.

The latent vectors are preprocessed by a Multilayer Perceptron (MLP) a stack of three blocks of Dense + Normalization + ReLU layers (Fig.~\ref{fig:fig_decoder}, bottom), concatenated together and passed to a 512-cell GRU layer. The output is postprocessed by another MLP and then passed to two dense layers that will provide the parameters for the synthesizers.

To synthesize audio, the DDSP library uses \textbf{Spectral Modelling Synthesis} (SMS), a technique proposed by Xavier Serra and Julius Smith \cite{serra1990spectral}, where the sound is modeled as two components: an additive or harmonic synthesizer, where a series of sinusoids is combined, and a subtractive synthesizer where the residual component is modeled as filtered noise. The DDSP library implements a differentiable version of the SMS, with an additional constraint: all the frequencies are integer multiples of f0. The expressiveness of this technique is a consequence of the high number of parameters needed. With the default configuration (60 harmonics, 65 noise magnitudes and 1 amplitude), sampled at 250Hz, 1 second of audio yields (60+65+1)*250 = 31,500 dimensions vs 16,000 audio samples.

The \textbf{additive (harmonic) synthesizer} models the harmonic part of the signal \(x(n)\) as a sum of \(K\) sinusoids, with different amplitudes \(A_{k}(n)\) and instantaneous phases \(\phi_{k}(n)\).

\begin{equation}
  \label{eq:harmonics}
  x(n) = \sum_{k=1}^{K} A_{k}(n)\sin(\phi_{k}(n))
\end{equation}

To improve interpretability, we can further expand \(A_{k}(n)\) into a global amplitude \(A(n)\) and a normalized distribution \(c(n)\) of amplitudes for the harmonic components, so \(A_{k}(n) = A(n)c_{k}(n)\).  Then Equation~(\ref{eq:harmonics}) can be rewritten as

\begin{equation}
x(n) = A(n)\sum_{k=1}^{K} c_{k}(n)\sin(\phi_{k}(n))
\end{equation}

The instantaneous phase --as the frequencies are multiples of f0-- is described as
\begin{equation}
  \label{eq:phase}
  \phi_{k}(n) = 2\pi\sum_{m=0}^{n} kf_{0}(m)+\phi_{0,k}
\end{equation} where \(\phi_{0,k}\) is the initial phase for the harmonic \(k\).

To reconstruct the additive signal at audio sample rate, f0 is upsampled using bilinear interpolation, and the amplitude and harmonic distribution are smoothed using an overlapping Hamming window centered on each frame.

The \textbf{subtractive synthesizer} models the residual component, the difference between the original signal and the signal from the additive synthesizer. If we assume that the residual component is stochastic, it can be modeled as filtered white noise, using a time-varying filter. The filter is applied in the frequency-domain, to avoid phase distortion.  For every frame \(l\), the network outputs \(H_{l}\), the frequency-domain transfer function of the FIR filter. The convolution is applied as a multiplication in the frequency-domain, and then an Inverse Discrete Fourier Transform (IDFT) is applied to recover the filtered signal.
\begin{equation}
y_{l} = IDTF(Y_{l}) \text{ where } Y_{l} = H_{l}DTF (x_{l}).
\end{equation}

Training the autoencoder means finding a set of parameters for the synthesizers that minimize the reconstruction loss i.e., minimize the difference between the output and input signals. A sample-wise loss measure is not recommended, as two similar waveforms can be perceived as having a very different sound. The spectral loss \(L\) --the difference between the spectrograms of the input (\(S\)) and output (\(\hat S\)) signals-- is perceptually better, but there is a compromise between time and frequency.

To solve this problem, a multi-scale spectral loss is defined. Instead
of using a single pair of spectrograms \((S, \hat S)\), the loss is
defined as the sum of different spectral losses \(L_{i}\) where \(i\) is
the FFT size. For this model, the loss is defined as

\begin{equation}
L = \sum_{i} L_{i}\text{ where } i \in (2048, 1024, 512, 256, 128, 64)
\end{equation}
and

\begin{equation}
L_{i} = || S_{i}-\hat S_{i}||_{1} + ||log( S_{i})- log(\hat S_{i})||_{1}
\end{equation}

\section{Experiments and results}\label{ch:ch4label}

This section describes the experiments carried out to understand the behavior, limitations and advantages of the model. The first experiment is a simple test to check the system is correctly set up. The second one trains several models using different datasets, including two datasets with multiple voices. The third one trains several models using the same dataset, but with different set of parameters for the spectral synthesizer. The last experiment introduces the latent space. 

To test the model, we will use a fragment of 'Over the rainbow' as sung
by Lamtharn (Hanoi) Hantrakul as the original audio presented to the
model. This melody is also used in the DDSP \cite{engel2020ddsp}
examples.

The notebooks and configuration files are available at \url{https://github.com/juanalonso/DDSP-singing-experiments}. The output of the experiments (audio and spectrograms) are available
online at \url{https://juanalonso.github.io/DDSP-singing-experiments/}.

\hypertarget{pre-evaluation}{%
\subsection{Pre-evaluation}\label{pre-evaluation}}

A quick test of the system has been carried out to reproduce the timbre
transfer experiments while checking the validity of our setup. In this
test we have used the same material as in the original paper (movements
II, III, IV, VI and VIII from the Bach Violin Partita no.1 BWM 1002, as
played by John Garner\footnote{\url{https://musopen.org/music/13574-violin-partita-no-1-bwv-1002/}})
generating the dataset and training the model for 30k steps using the
standard configuration file {\texttt{solo\_instrument.gin}} (60 sine
waves, 65 noise magnitudes, reverb and batch size of 16).

\begin{figure}
\centering
\includegraphics[width=0.99\linewidth]{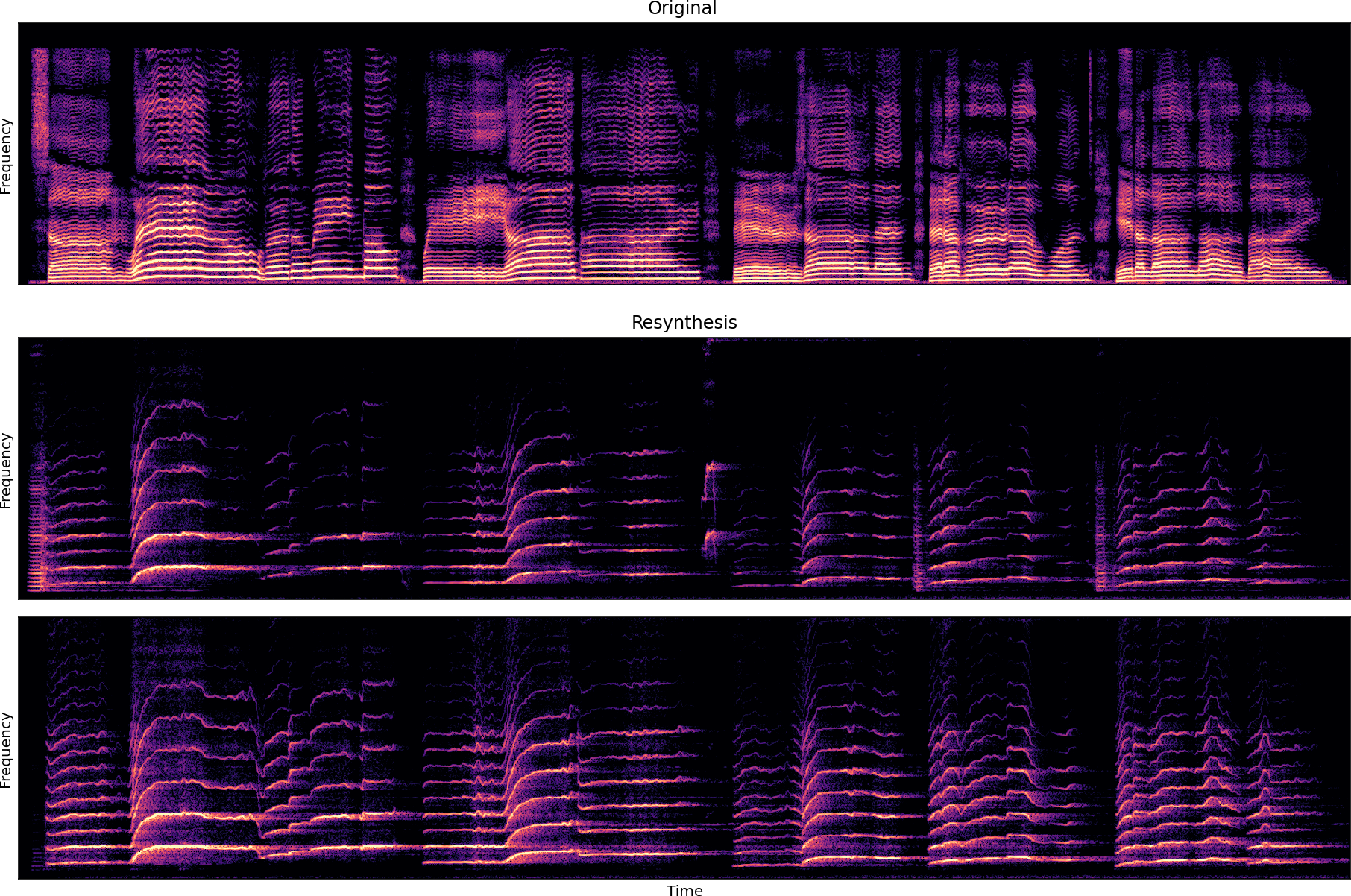}
\caption{Over the rainbow as sung by Lamtharn (Hanoi) Hantrakul (top)
and reconstructed by our version of the violin model (+2oct, -20dB),
using no statistical information (middle) and with statistical
information (bottom).}\label{fig:fig_preeval}
\end{figure}

The estimated f0 is transposed up two octaves, to better match the pitch
range of the violin, and the loudness is attenuated 20dB to match the
model amplitude. Two reconstructions are produced, the first one with no
additional information, and the second one using masking and statistical
amplitude adjustment. These reconstructions are shown
in Fig.~\ref{fig:fig_preeval}.

The violin model works as expected. The resulting audio is equivalent to
the audio produced by the original model, which is trained for 20k extra
steps with a batch size of 32.

\hypertarget{different-datasets-trained-on-the-same-parameters}{%
\subsection{Different datasets trained on the same
parameters}\label{different-datasets-trained-on-the-same-parameters}}

\subsubsection{Single voice model}\label{ch:svm}

To generate the models for singing voices, we are using seven datasets (see Table \ref{tab:datasets}), obtained from speakers and singers, both male and female, in different languages.
No audio source separation software is
used. Two of the datasets were scrapped from YouTube\footnote{\url{https://www.youtube.com/channel/UCyJWqwbI7CPAZpCqJ_yTg2w}}
(mrallsop) and the Alba Learning website\footnote{\url{https://albalearning.com/}}
(alba). The rest of the vocal tracks (eva, belen, servando) are raw
vocal tracks recorded by professional singers and have been kindly
provided by their interpreters upon request of the authors. These tracks
were not specifically recorded for this work, they were previously
recorded for existing and upcoming music records. Two additional
datasets have been created by combining the audio files from the eva and
belen datasets (voices2) and the eva, belen and servando datasets
(voices3). Files longer than five minutes (mrallsop and alba) have been
split into three to four-minute chunks. Other than that, the original
audio has not been transformed in any way, keeping all the original
characteristics: silences, different volume levels, background noise,
etc.

\begin{table}[]
    \begin{center}
        \begin{tabular}{llll}
            \textbf{Name}     & \textbf{Gender} & \textbf{Type}& \textbf{Loss}          \\
            servando & male   & sung lyrics &   6.984\\
            belen    & female & sung lyrics &   5.114\\
            eva      & female & sung lyrics &   4.987\\
            mrallsop & male   & spoken speech & 5.437\\
            alba     & female & spoken speech & 5.142\\ 
            voices2   & female  & sung lyrics &    5.415 \\
            voices3   & mixed  & sung lyrics &    6.143 \\ 
        \end{tabular}
          \end{center}
    \caption{List of datasets used for training the model and loss value after 40k steps.}
        \label{tab:datasets}

\end{table}

The models are trained for 40k steps each, using the notebook
{\texttt{01\_train}}. Each model has been trained using {\texttt{singing.gin}}, a modified version of the standard configuration file.

\begin{figure}
\centering
\includegraphics[width=\linewidth]{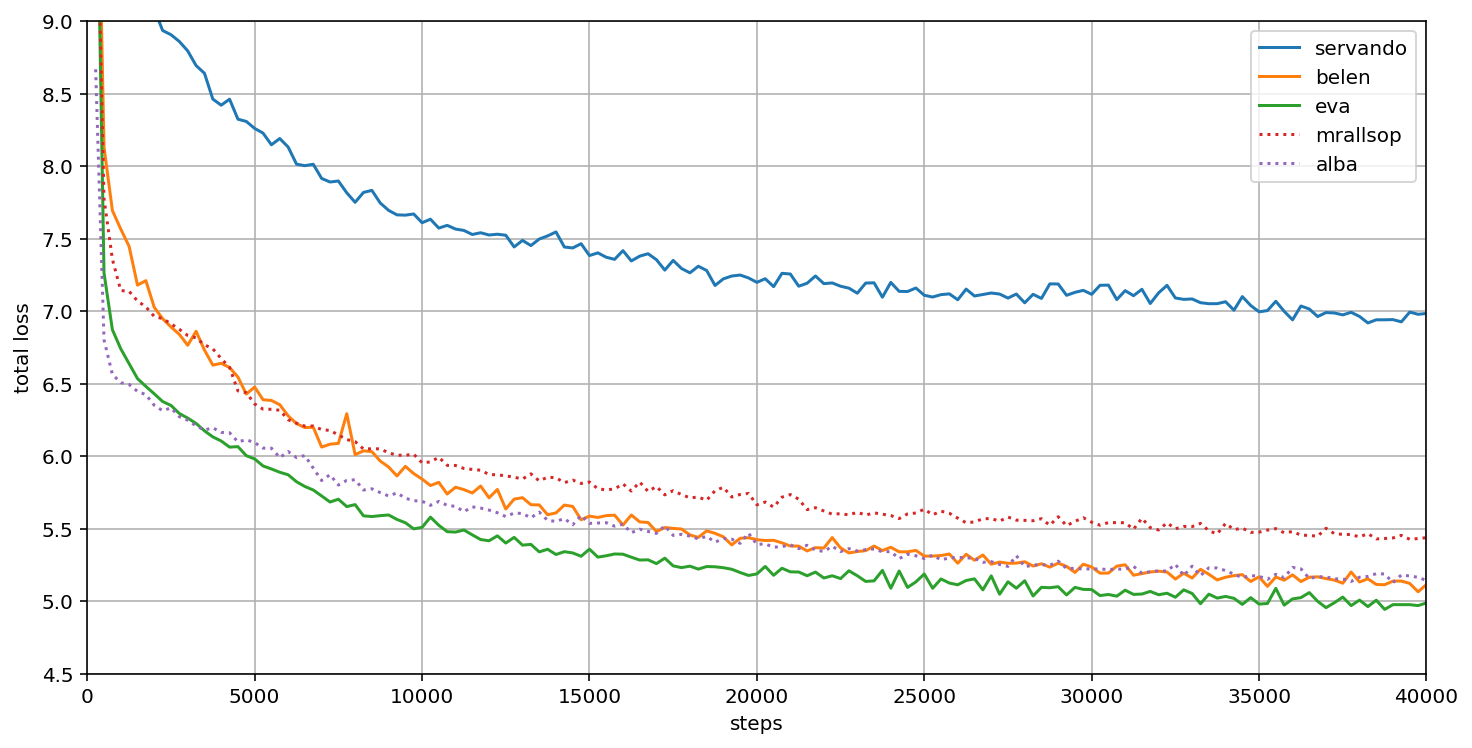}
\caption{Spectral loss for the five independent datasets. The dotted
lines represent the two spoken speech
datasets.}\label{fig:fig_loss_voices}
\end{figure}

The losses after training (Fig.~\ref{fig:fig_loss_voices}) are all in the range
{[}4.987, 5.437{]} as recommended in \cite{engel2020ddsp}. There are no
significant differences between the losses in the spoken and sung
datasets. The servando model, whose loss is considerably higher, is an
exception. The only apparent difference with the rest of datasets is
that servando's source audio is hard-clipped / compressed at 0dB, with a
smaller dynamic range than the other voices, which present a wider range
of amplitudes and compression values.

The output is generated by executing notebook {\texttt{02\_run}}. The
pitch shift is chosen manually by comparing the mean MIDI pitch of the
dataset with the mean MIDI pitch of the melody. The loudness shift is
handpicked after comparing different settings. The threshold and quiet
parameters are adjusted manually depending on how much noise bleeds into
the silent parts of the original audio. The values chosen for each
example are shown in the audio web page.

\hypertarget{multiple-voices-model}{%
\subsubsection{Multiple voices model}\label{multiple-voices-model}}

\begin{figure}
  \centering \includegraphics[width=\linewidth]{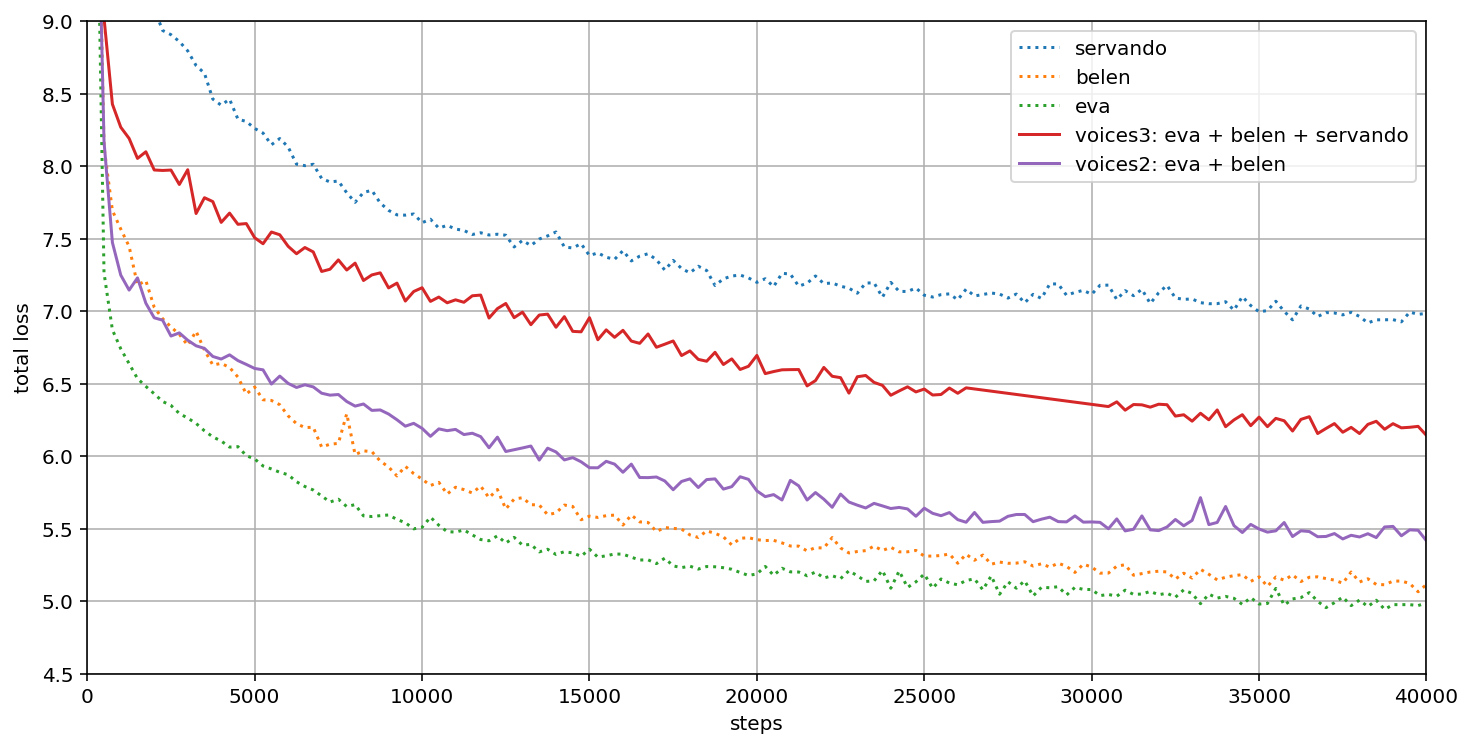}
  \caption{Spectral loss for the hybrid dataset (voices2 and voices3, shown as bolder lines), compared with the other sung lyrics datasets.  The flat segment between 25k and 30k steps in voices3 is due to a problem with the logged data.\label{fig:fig_loss_combined}}
\end{figure}

The model \textbf{voices2} combines the source audio from belen and eva,
both female singers. It is trained for 40k steps. The loss, after training is 5.415, higher than the loss of both of the
datasets (belen=5.114, eva=4.987), as shown in Fig.~\ref{fig:fig_loss_combined}. This result is expected,
as we are training the model with the same number of steps as the single
voice models but, in this case, the model needs to learn two different
timbres, with different loudness and slightly different MIDI mean
pitches (belen=62.39, eva=64.10).

When the model is used, the resulting audio is a combination of the individual voices. Depending on the f0 and loudness, the model outputs a succession of fragments by the original singers. The transition is smooth and there is no chorus effect: only one voice is perceived at a concrete frame.

The model \textbf{voices3} combines the source audio from belen, eva and
servando. It is trained for 40k steps: %
the model must learn three timbres, and one of them is
more different from the other two (servando, loss=6.984, MIDI mean
pitch=55.08) The peculiarities of the servando dataset penalize the
training, and thus the model presents a higher loss than voices2.

Fig.~\ref{fig:fig_loss_combined} shows that the loss
of voices3 (6.143) is lower than the loss of the servando model (6.984).
We attribute this effect to an imbalanced dataset: the duration of
servando's source audio is 11 minutes and 02 seconds, whereas belen's
and eva's source audio combined is 28 minutes and 50 seconds.

In this experiment, the voice mixing capabilities of the model are more
pronounced than in the previous experiment. The mean MIDI pitch of the
example song is 51.30. Considering that the mean MIDI pitches of the
servando, belen and eva datasets on Table \ref{tab:datasets} are, respectively 55.08,
62.39 and 64.10, we can expect that when rendering the audio, the model
will generate a mix of the nearest-pitched voices. This is the case:
when using the example fragment without transposing, the resulting
melody is a mix of servando's and belen's voice. If the example fragment
is transposed an octave higher (MIDI pitch of 63.30) the resulting
melody is a mix of belen's and eva's voice. To demonstrate this effect,
six examples have been uploaded to the audio page, using different sets
of preprocessing parameters.

\hypertarget{single-dataset-different-spectral-parameters}{%
\subsection{Single dataset, different spectral
parameters}\label{single-dataset-different-spectral-parameters}}

In this experiment, we want to understand how the parameters of the
spectral synthesizer affect the learning process and the results. We
will be using the eva dataset --since its model has the lowest loss
after training-- to train the model using a range of spectral
configurations.

We chose three values for the harmonic component (20, 60 and 100
sinusoidal waves). 60 is the default value provided in the configuration
files. 100 is the maximum value a model can be trained without getting
out of memory errors, and 20 is the symmetrical lower bound
(\(60-(100-60)\)). For the noise component we chose 10, 35 and 65 noise
magnitudes, 65 being the default value.

Nine models are generated, one for each combination of harmonic
components and noise magnitudes. Each model is trained for 20k steps,
using the same configuration file we used in the previous sections
({\texttt{singing.gin}}). f0 and loudness are preprocessed using the
handpicked parameters from Section \ref{ch:svm} and
shown in Table \ref{tab:preprocessing_parameters}.

\begin{table}[]
  \begin{center}
    \begin{tabular}{c|c}
      \textbf{Use statistics} & Yes \\
      \textbf{Mask threshold} & 1\\
      \textbf{Quiet} & 20 \\
      \textbf{Autotune} & 0 \\
      \textbf{Octave shift} & +1\\
      \textbf{Loudness shift} & -10dB
    \end{tabular}
  \end{center}
  \caption{f0 and loudness preprocessing parameters.\\ \label{tab:preprocessing_parameters}}\tabularnewline
\end{table}

\begin{figure}
\hypertarget{fig:fig_loss_eva}{%
\centering
\includegraphics[width=\linewidth]{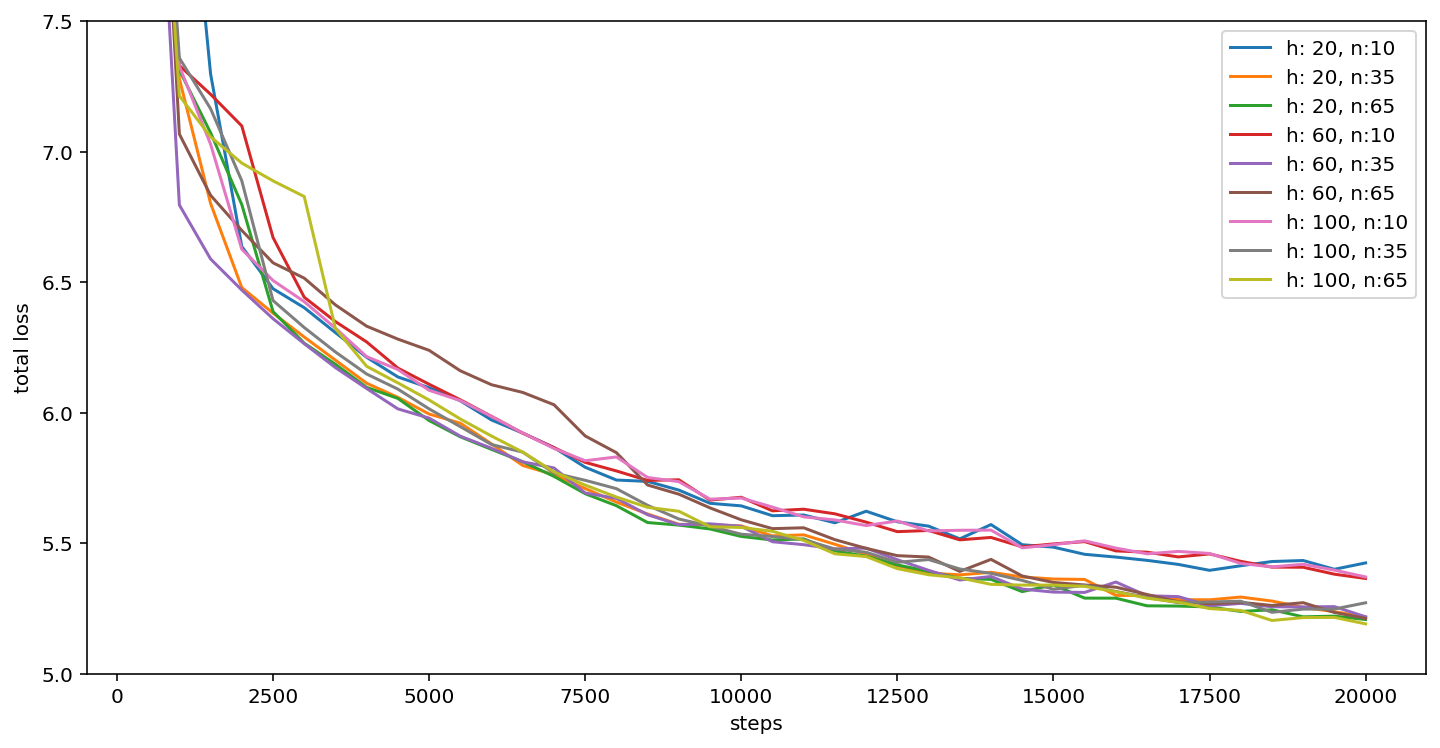}
\caption{Total (spectral) loss for the eva dataset, using different
parameters for the spectral synthesizer. h is the number of harmonic
components; n is the number of noise
magnitudes.}\label{fig:fig_loss_eva}
}
\end{figure}

As can be observed in Fig.~\ref{fig:fig_loss_eva}, to decrease the spectral loss, the amount of noise magnitudes is more relevant than the number of harmonic components. Perceptually, more models and tests are needed. On an informal test where three users with no musical training were presented pairs of the snippet 'way up high' (seconds 7 to 11 in the original audio) rendered with different parameters, there was no agreement on which "sounded better". The only exception was snippet h:20
n:10, where the subjects remarked it was the worst sounding of the pair. All subjects commented on listening fatigue due to being exposed to the same melody.

\begin{figure}
\centering
\includegraphics[width=0.99\linewidth]{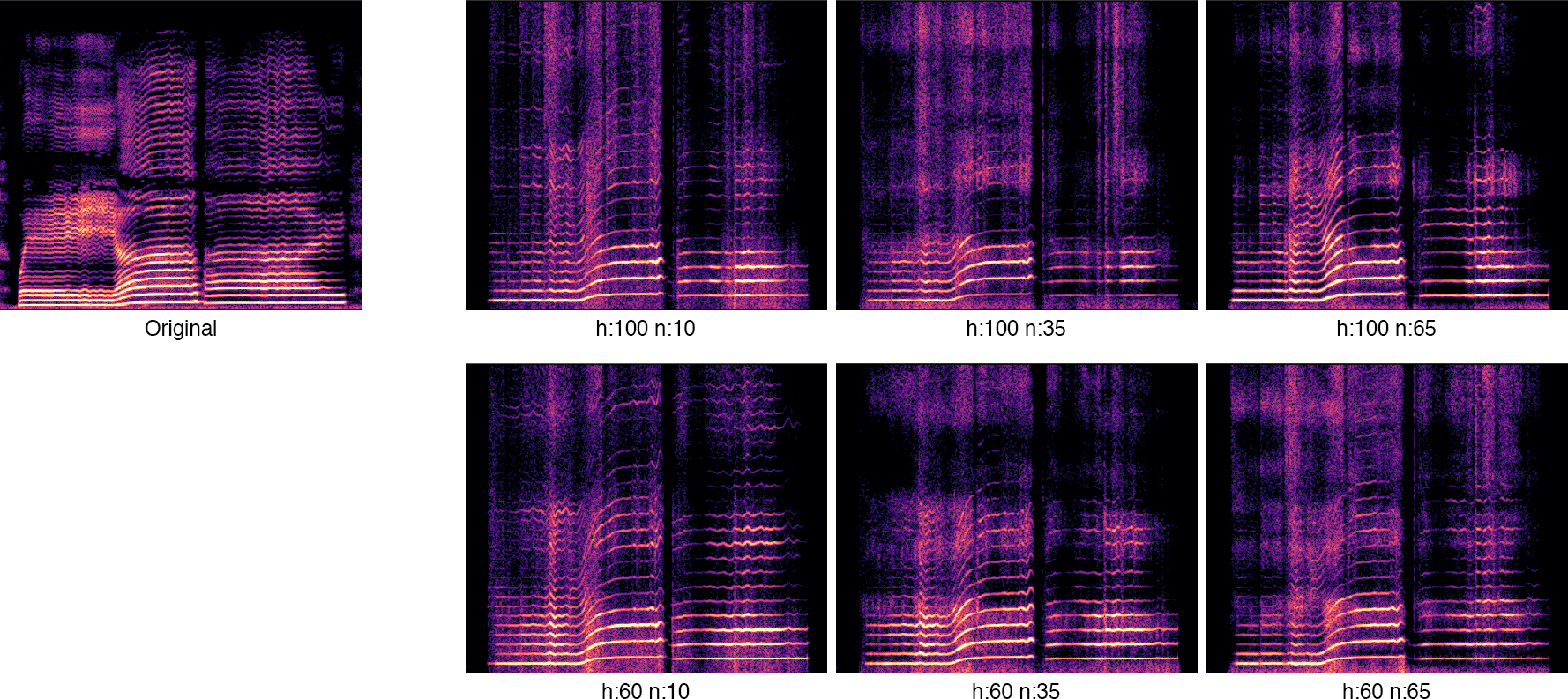}
\caption{Spectrogram of the original audio ('way up high', seconds
7-11). left column, and spectrograms of the model's output with
different spectral parameters.}\label{fig:fig_spectrogram_spectral}
\end{figure}

Observing the spectrograms of the reconstructions in Fig.~\ref{fig:fig_spectrogram_spectral}, the examples with
the lowest number of noise magnitudes (\(n=10\)) show the model trying
to reconstruct the high frequency noise with the harmonic model (faint
sinusoids in the spectrograms).

\hypertarget{adding-the-latent-space}{%
\subsection{Adding the latent space}\label{adding-the-latent-space}}

\begin{figure}
  \hypertarget{fig:fig_timbre_transfer_Z_architecture}{%
    \centering \includegraphics[width=\columnwidth]{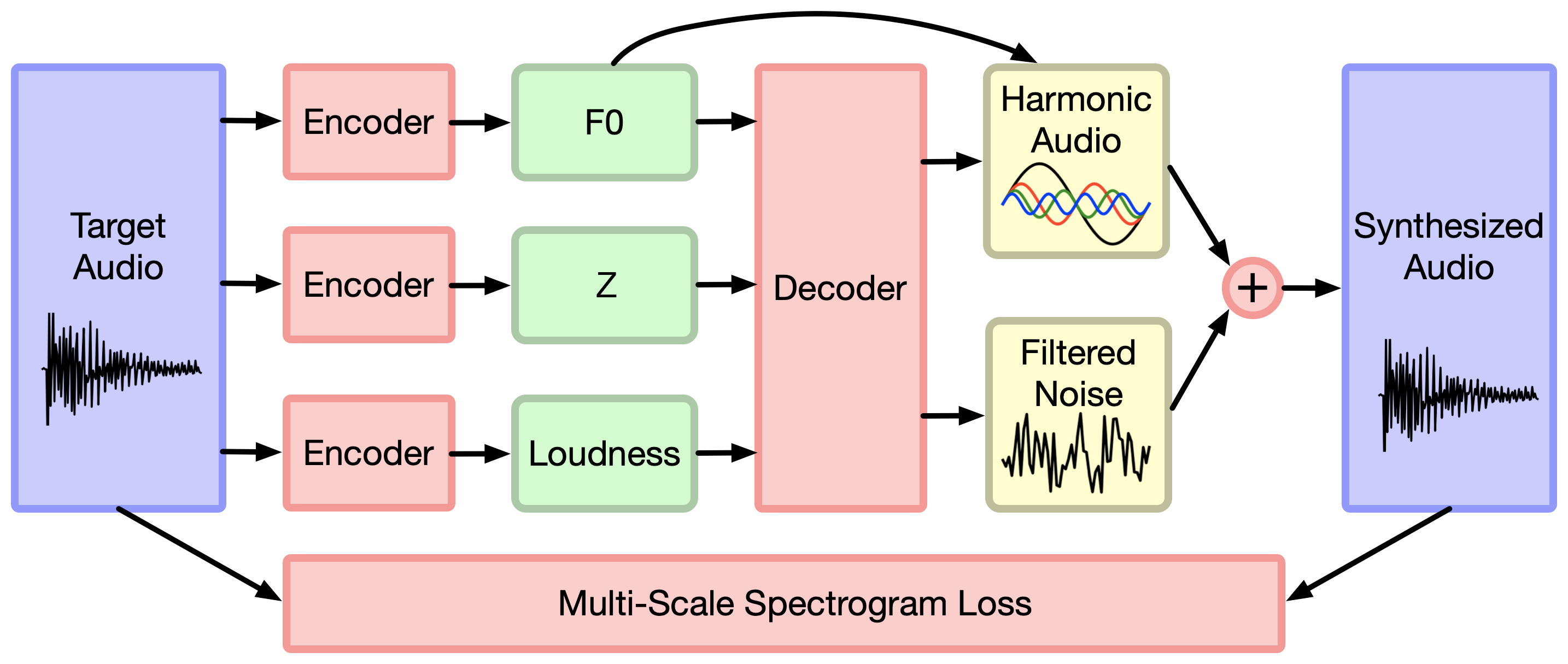}
    \caption{Timbre transfer architecture with latent space. Adapted from
\cite{engel2020ddsp}.}\label{fig:fig_timbre_transfer_Z_architecture}
}
\end{figure}

For our last experiment, we expand the {\texttt{singing.gin}} configuration (Fig.~\ref{fig:fig_timbre_transfer_Z_architecture}) to include a time-varying encoding of the Mel Frequency Cepstral Coefficients (MFCC) to understand how this additional encoding affects the model's output. This representation is specially well suited for the human voice, as it mimics the non-linearity of human sound perception and helps model the timbre \cite{godino2006dimensionality}. The MFCC are encoded using a normalization layer and a 512-unit GRU, reducing the dimensionality from 30 coefficients to a 16-dimensional latent vector, at 125 frames per second and then upsampled to 250 frames, to match the sampling rate of the f0 and loudness vectors. To decode this vector, the same MLP architecture shown in Fig.~\ref{fig:fig_decoder}, bottom, is used, and concatenated with the preprocessed f0 and loudness vectors.

To generate the z models, we used the belen and mrallsop datasets and a new dataset, felipe\_vi, extracted from Felipe VI's, King of Spain, 2020 Christmas speech (male spoken voice, Spanish) As with the previous datasets, felipe\_vi is used "as is", without any kind of preprocessing. A new configuration file ({\texttt{singing\_z.gin}}) is also used, and it is available at the GitHub repository. This configuration file inherits all the values from {\texttt{singing.gin}} and includes the z encoder.

\begin{figure}
\centering
\includegraphics[width=\columnwidth]{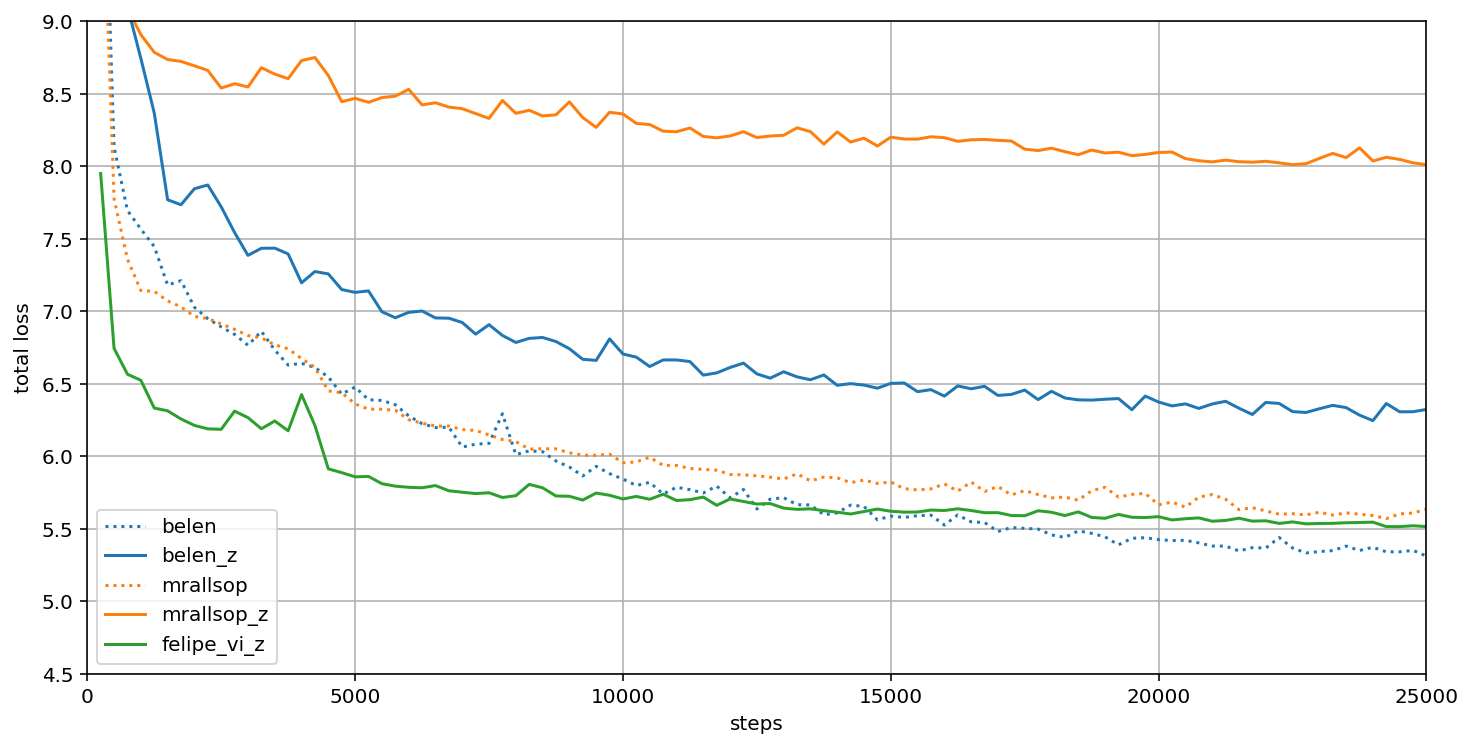}
\caption{Spectral loss for the three z models. The dotted lines are included for reference, and represent loss function for the two related non-z models.}\label{fig:fig_loss_z}
\end{figure}

The models are trained for 25k steps. As shown in Figure \ref{fig:fig_loss_z}, with the additional encoding, the loss function increases (from 5.310 to 6.322 in the case of the belen dataset and from 5.634 to 8.011 in the mrallsop dataset) This is an expected behavior, as the model now needs to fit additional parameters. The loss value for the felipe\_vi\_z model is low enough (5.514) to be in the range of the non-z models.

\section{Discussion}\label{ch:ch5label}

\subsection{Results}\label{results}

The architecture proposed in \cite{engel2020ddsp} is powerful enough to
perform timbre transfer and to produce surprising results even if the
dataset is small (10-15 minutes of audio). The training, using the Colab
environment, is fast and it works out of the box, with the GPU
infrastructure ready. All the required libraries (CREPE, TensorFlow,
Apache Beam...) are available without version conflicts.

Regarding singing voice synthesis, we did not make it easy for the model
(small datasets, unprocessed audio...), but the quality of the results
surprised us. Of course, in no way the output of the model is going to
be mistaken for a human, but the model's ability to produce such good
results with no additional conditioning is very promising and opens
several avenues for exploration, research and creative usage.

With the addition of the z-encoder, the quality of produced audio is increased, becoming almost intelligible. The right vowels start appearing, and the babbling effect is reduced substantially. The resulting timbre is halfway between the instrument timbre and the original timbre.

This architecture makes very difficult to estimate the performance of a model. As we have noticed, the training loss of the servando model is quite high, compared with the rest, but when analyzing the dataset, nothing stands out as the cause of this value. Similar datasets (speaking, male voice) such as felipe\_vi\_z and mrallsop\_z present very different losses (5.514 versus 8.011 respectively), but the quality of the resulting audio is comparable.

\subsection{Problems}\label{problems}

\subsubsection{Babbling and stuttering}\label{ch:babbandstutt}

We are forcing the model to recreate sung lyrics. The model needs to
learn the timbre of the voice, how to extrapolate previously unheard
pitches and the flow of the language. The current architecture manages
to extract the timbre and to correlate f0 and loudness with sounds, 
but it lacks the ability to learn the sequences of phonemes that constitute
speech. Even with more comprehensive datasets, where all the possible
combinations of phonemes and pitches could be present, without
additional conditioning (phonetic, text, ...) the model will try to make
up what to say, and the produced audio will be just a better-quality
version of this stuttering and nonsensical babbling.

During the development of this work, the audio has been presented to
several listeners who knew the original singers (belen, servando) and
they all found the results unsettling, due to this kind of babbling.
They recognized the speaker, but the babbling and stuttering were
compared to listening to a person having suffered a stroke that impaired
the language centers of the brain.

\hypertarget{silence}{%
\subsubsection{Silence}\label{silence}}

If the dataset does not include silences (for example, the dataset used
to train the violin model) the resulting model will have some
difficulties trying to recreate them and will resort to generate some
very low notes. This can be mitigated by adding some transitions to and
from silence and by fine tuning the preprocessing parameters, which
right now it is a manual process dependent on the input audio and the
model.

The example on the audio page shows this phenomenon particularly well.
On the one hand, the original audio, a staccato synthesizer riff, is
played as legato. On the other hand, the silence that occurs at seconds
3 and 9 is reinterpreted as a pair of low-pitched tones. Even tweaking
the preprocessing parameters, we can mitigate the low tones, but not
suppress them.

\hypertarget{pitch-artifacts}{%
\subsubsection{Pitch artifacts}\label{pitch-artifacts}}

The accuracy of the output pitches depends on the f0 estimation. If the
estimation made by the CREPE model is wrong, the output will include
wrong notes. We have detected several cases where CREPE estimates a very
high pitch with high confidence, so the preprocessor cannot mask it. In
those cases, the resulting audio will include a squeal, a quick
glissando to the estimated high pitch.

To avoid this, we can substitute CREPE for another algorithm, or use a
symbolic notation, such as MIDI, to generate the f0 vector. In that
case, we risk having a very robotic monotonous voice, and we would need
to add some modulation (amplitude, pitch...) to make it more natural
sounding.

\hypertarget{ddsp-maturity}{%
\subsubsection{DDSP maturity}\label{ddsp-maturity}}

Although the possibilities of these libraries are immense, exploring
them is a challenging task for two major reasons. The first one is the
lack of information about the most complex processes, how some of its
modules work and how to modify the workflow to add new functionalities.
Despite open-sourcing the code in GitHub, the tutorials and demos barely
scratch the surface. The second problem is that, because the libraries
are still under active development, some of the updates are missing
information about the release changes and are not as stable as expected.

\hypertarget{ch:ch6label}{%
\section{Conclusions and Future Work}\label{ch:ch6label}}

\hypertarget{conclusions}{%
\subsection{Conclusions}\label{conclusions}}

The DDSP library opens up a lot of possibilities for audio synthesis.
The work presented here allows us to get a better understanding on how
the DDSP library works, especially when used for timbre transfer. It
achieves two goals:

\begin{enumerate}
\item
  \textbf{Test the validity of the DDSP architecture to generate a
  singing voice.} The tests carried out on the architecture have used
  unfavorable data (no preprocessing, background noises, small datasets,
  etc.), and even so, the network has been able to generate audio
  similar to the human voice, with enough features to be recognized as
  belonging to a specific singer.
\item
  \textbf{Create an easy-to-use environment to facilitate model training
  and timbre transfer to end users}. The notebooks provided will help the SMC community to ease the learning curve of this architecture and get
  familiar with the advantages and nuisances of the library. Since the
  GitHub repository includes some of the models used in this work,
  curious users can just interact with them, without needing to create
  their own datasets. Also, as per today (March 2021) our models are
  compatible with the ones made available by Google (flute, violin...)
  so they can be used interchangeably.
\end{enumerate}

\hypertarget{future-work}{%
\subsection{Future Work}\label{future-work}}

In order to create a more refined model which is capable of synthesizing
much realistic utterances with lyrics replication (and thus avoiding the
gibberish / stuttering effect) additional work must be done in the
following areas:

\hypertarget{conditioning}{%
\subsubsection{Conditioning}\label{conditioning}}

As noted in Section \ref{ch:babbandstutt}, the phonetic output is made up by the model, without any reference to real speech. The current architecture does not condition the output in any other data than pitch and loudness, missing a lot of additional information present in sung lyrics. To get the nuances of human singing and model the lyrics, we need to include additional conditioning on the language level, for example, the phonetic conditioning proposed in \cite{hutter2020timbre}.

\hypertarget{use-representations-more-suitable-to-voice-synthesis}{%
\subsubsection{Use representations more suitable to voice
synthesis}\label{use-representations-more-suitable-to-voice-synthesis}}

The default architecture proposed in the DDSP is generic, designed for monophonic musical instruments. Using mel-spectrograms as proposed in \cite{fabbro2020speech}, instead of using raw audio or by postprocessing the harmonic and the noise components of the transformed audio to balance the voiced and unvoiced parts of the speech \cite{hutter2020timbre}, results could be improved.

\hypertarget{use-synthesizers-more suitable-to-voice-modelling}{%
\subsubsection{Use synthesizers more suitable to voice modelling}\label{use-synthesizers-more suitable-to-voice-modelling}}
As stated previously, by using Spectral Modelling Synthesis, we get a very expressive synthesizer at the expense of producing twice as much data per second as the sampled audio. However, other synthesizers can provide a more compact representation, resulting in a smaller model which will be faster to train and run. The authors are currently working on implementing both an AM and a 2-operator FM differentiable synthesizer. These simple synthesizers will provide us a better understanding of the capabilities and nuances of a differentiable synth, its performance, how to integrate them in the existing toolchain, and how to modify the model architecture to fit different synthesizers.

\hypertarget{preprocessing-of-f0}{%
\subsubsection{Preprocessing of f0}\label{preprocessing-of-f0}}

Even if the model is able to transfer the timbre perfectly, following the "Garbage in, garbage out" concept, the quality of the output will be affected by the quality of the latent vectors. If the pitch estimation is not accurate, the resulting audio will present pitch artifacts. A quick solution can be to extract f0 from MIDI. While the resulting f0 is going to be precise, it is going to lack expressiveness. Solutions as the one proposed in \cite{jonason2020control} can add expressiveness to the MIDI data.

\hypertarget{explore-the-creative-possibilities-of-the-model}{%
\subsubsection{Explore the creative possibilities of the
model}\label{explore-the-creative-possibilities-of-the-model}}

The creative possibilities offered by the DDSP architecture are immense, either with low fidelity, glitchy results as explored in this work, or with more realistic results by applying additional conditioning. Some of the possibilities are pitch- and time-shifting, lyric translation, voice morphing, change of singing style (e.g., to and from opera, pop, blues), tremolo and vibrato removal or addition, to name just a few.

\bibliography{bib/mybib} 

\end{document}